\documentstyle[12pt,epsfig]{article}
\begin{document}
\begin{center}
{\bf PHOTOMETRIC OBSERVATIONS\\
 AND\\
 APSIDAL MOTION STUDY OF V1143 CYGNI}\\ \ \\
{\small {\bf A. DARIUSH}\\
{\it Physics Department, Azad University of Arsenjan, Arsenjan, Iran\\
IUCAA, Post Bag 4, Ganeshkhind, Pune-411 007, Pune, India; \\ E-mail: dariush@iucaa.ernet.in}\\ \ \\

{\bf N. RIAZI}\\
{\it Physics Department and Biruni Observatory, Shiraz University, Shiraz 71454, Iran\\
Institute for Studies in Theoretical Physics and Mathematics (IPM), Farmanieh, Tehran, Iran; E-mail: riazi@physics.susc.ac.ir}\\ \ \\

{\bf A. AFROOZEH}\\
{\it Biruni Observatory, Shiraz University, Shiraz 71454, Iran }}\\ \ \\

\end{center}

\begin{center}
{\bf  Abstract}
\end{center}

 {\small Photometric observations of the eccentric eclipsing binary
V1143 Cygni were performed during the August-September 2000 and
July 2002, in B and V bands of the Johnson system.  The analysis
on both light curves was done separately using the 1998 version of
Wilson's LC code. In order to find a new observed rate of apsidal
motion, we followed the procedure described by Guinan and Maloney
(1985). A new
 observed rate of apsidal motion of 3.72
degrees/100yr was computed which is close to the one reported
earlier by Khaliullin (1983), Gimenez and Margrave (1985), and
Burns et al.(1996). }
\section{{\bf Introduction}}
V1143 Cygni (HD 185912; HR 7484; BD$+54^\circ 2193$;
$V_{max}=+5.86$; B-V=+0.46;
 $\alpha=19^h 38^m 41^s.18$;  $\delta=+54^{\circ} 58'25''.7$) is a
double-lined eclipsing binary, consists of a pair of F5V stars
with high orbital eccentricity (e=0.540) and a relatively long
period of 7.640 days (see Fig.1). According to Andersen et al.
(1987) who have determined the stellar and orbital properties of
this system  accurately, the corresponding radii and masses are
$r_1=1.346\pm0.023R_{\odot}$, $r_2=1.323\pm0.023R_{\odot}$ and
$m_1=1.391\pm0.016M_{\odot}$, $m_2=1.347\pm0.013M_{\odot}$,
 respectively.

\begin{figure}[h]
  \epsfxsize=6cm
  \centerline{
\epsffile{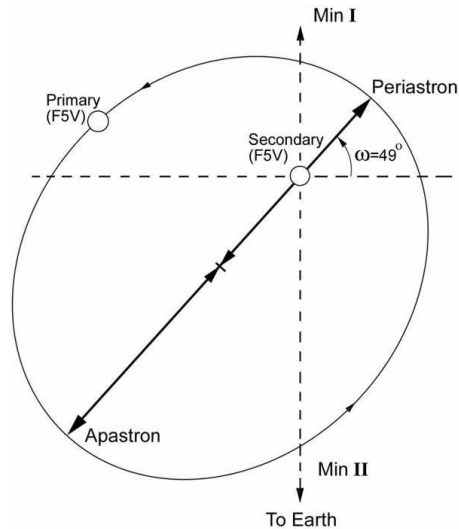} } \caption{Relative orbit of V1143 Cygni drawn
in scale.}
\end{figure}

It is known that V1143 Cyg is one of the best examples of
eclipsing binaries with apsidal motion, in which the observed rate
of apsidal motion is grater than the value  predicted by general
relativity and stellar evolutionary models.

The observed rate of apsidal motion is due to the contribution of
two terms; a classical term as well as the general-relativistic
term. In this sence, the observational apsidal motion rate is
\begin{equation}
\dot{\omega}_{obs}=\dot{\omega}_{cl}+\dot{\omega}_{GR},
\end{equation}\\
where $\dot{\omega}_{cl}$ denotes the classical or Newtonian term
and $\dot{\omega}_{GR}$ is the relativistic contribution which can
be determined using the formulas given by Gimenez (1985). In the
case of V1143 Cyg, the observed rate of apsidal motion is
$\dot{\omega}_{obs}=3^{\circ}.52/100^{yr}\pm
0^{\circ}.72/100^{yr}$ (Burns at al. 1996) while Andersen at al.
(1987) calculated a  faster theoretical apsidal motion of
$\dot{\omega}_{theo}=4^{\circ}.25/100^{yr}\pm
0^{\circ}.72/100^{yr}$ in which the expected relativistic and
classical (Newtonian) contributions to apsidal motion are
$\dot{\omega}_{cl}=2^{\circ}.39/100^{yr}$ and
$\dot{\omega}_{GR}=1^{\circ}.86/100^{yr}$, respectively. It is
seen that the classical and relativistic contributions are of the
same order in this system.

\section{{\bf Observations}}
V1143 Cygni was observed during 24 nights from July to September
2000 at Biruni Observatory of Shiraz University
 (Longitude: $52^{\circ}31'$ E, Latitude: $29^{\circ} 36'$ N).
Observations were made with a 51cm  cassegrainian telescope
equipped with an uncooled RCA4509 multiplier phototube. Two stars
HD 184240 ($V_{max}=+6.30$) and HD 186239 ($V_{max}=+7.10$)
 were selected as comparison and
check stars respectively. The integration time for all of the
observations were fixed to 10 seconds. The output signals of the
photomultiplier were fed to a computer after amplification, using
an A/D convertor. The measurements were made using B and V filters
of intermediate-bandpass blue $\lambda_{max}=4400 \AA$ and yellow
$\lambda_{max}=5530 \AA $ which are matched closely to the
Johnson's UBV system. Times were converted to Heliocentric Julian
Day Number (HJD). Data reduction and atmospheric corrections were
done to obtain the complete light-curves in two filters, using a
computer code developed by G. P. McCook. Figures 2 and 3,
represent the observed light curves in B and V filters
respectively. In order to enhance the accuracy of the present work
, two other minima were observed on July 16 and 18, 2002. This
time, the star HD 185978 (F8; $m_V=+7^m.8$) was selected as the
comparison star.

\begin{figure}[h]
  \epsfxsize=8cm
  \centerline{
\epsffile{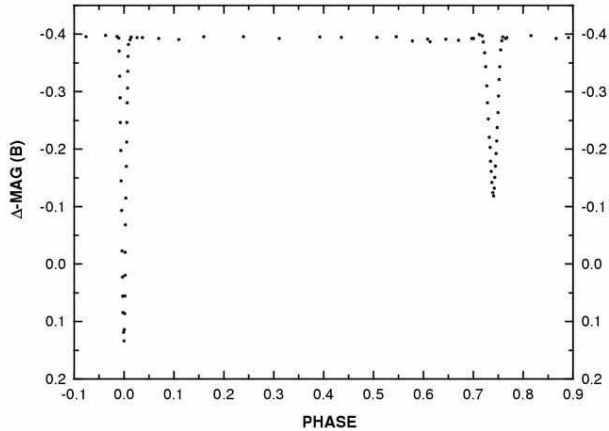} } \caption{Observed light curve of V1143 Cygni
in the B filter.}
\end{figure}

\begin{figure}[h]
  \epsfxsize=8cm
  \centerline{
\epsffile{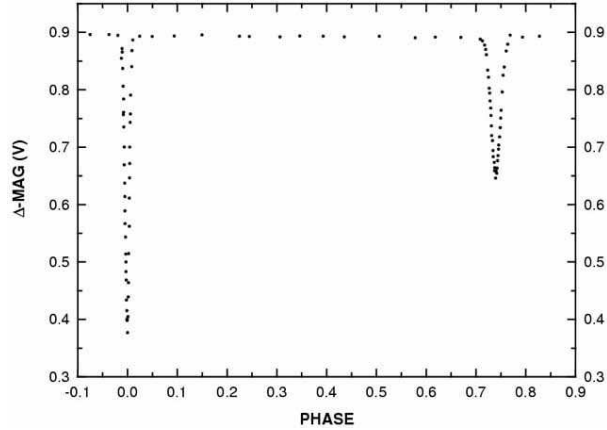} } \caption{Observed light curve of V1143 Cygni
in the V filter.}
\end{figure}


\section{{\bf Times of minima and light curve analysis}}

From the observed light curves, heliocentric times of minima  (one primary and one secondary)
were computed by fitting a Lorentzian function to the observed minima data points
(see Figures 4 and 5).
This function can be expressed as
\begin{equation}
y={y_o}+{\frac{2A}{\pi}} {\frac{w}{4(x-{x_c})^{2}+w^2}},
\end{equation}
where $y_o$ is the baseline offset, A is total area under the
curve from baseline, $x_c$ is the center of the minimum and $w$ is
full width of the minimum at half height. The minima were
calculated according to the ephemeris given by Andersen et al.
(1987)
\begin{equation}
Min.I=HJD 2449234.6144+7^{d}.64075217 \times E,\\
\end{equation}
and are given in Tables 2 and 3.
 The corresponding errors in primary and secondary minima are:
$\pm 0.00066$ and $\pm 0.00220$, respectively.
 Meanwhile, in each filter, the depths of minima are as follows:\\ \ \\
  Filter B: Min.I: $0^m.53\pm 0.02$, Min.II: $0^m.25\pm 0.02$\\ \ \\
    Filter V: Min.I: $0^m.48\pm 0.02$, Min.II: $0^m.23\pm 0.02$\\

\begin{figure}[h]
  \epsfxsize=8cm
  \centerline{
\epsffile{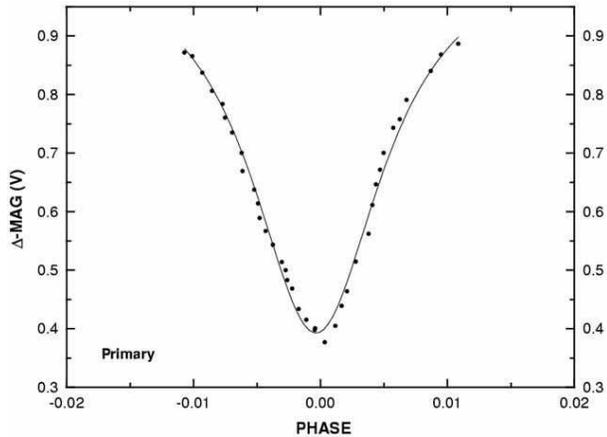} } \caption{A sample Lorentzian fit to the
primary minimum.}
\end{figure}

\begin{figure}[h]
  \epsfxsize=8cm
  \centerline{
\epsffile{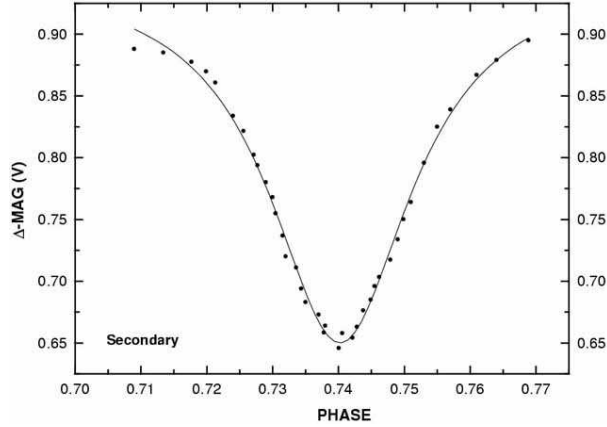} } \caption{A sample Lorentzian fit to the
secondary minimum.}
\end{figure}

The B and V light curves of V1143 Cyg  have been analyzed
separately by using the  Wilson code in order to derive
photometric elements of this system.
 The program consists of two main FORTRAN programs {\bf LC}
(for generating light and radial velocity curves) and {\bf DC} (to
perform differential corrections and parameter adjustment of the
LC output). The model which upon the program is based on, has been
described and quantified in papers by Wilson (1979, 1990, 1993).
Since the system V1143 Cygni is detached with both components
residing well inside their respective Roche lobes (Burns et al.
1996), the solution was performed in {\bf mode 2}. In our
analysis, we assumed a value of zero for the third light
($l_3=0$). Also we fixed the ratio of the axial rotation rate to
the mean orbital rate for stars 1 and 2 ($F_1,F_2$). Before
running the {\bf LC} code, the position of the preastron was
estimated from the apsidal motion study as  discussed in section
4. In order to optimize the observed parameters given in Table 1,
we wrote an auxiliary computer programm  to compute the sum of
squares of the residuals in both filters by using the{\bf LC}
output. Figures 6 and 7 show
 the sum of the squared residuals versus selected parameters for the two filters.
For example, Figure 6(a) represents the weighted sum of the
squared residuals $\sum\omega r^2$ (SSR) versus {\it eccentricity
(e)} and
{\it inclination angle (i)} for filter B.\\

\begin{figure}[h]
  \epsfxsize=11cm
  \centerline{
\epsffile{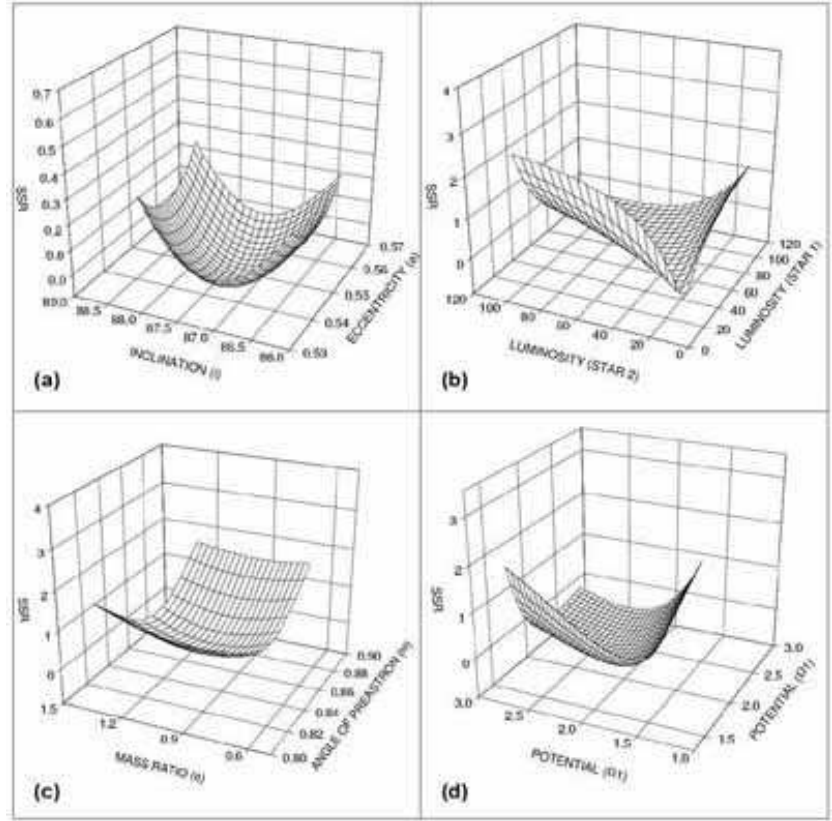} } \caption{The weighted sum of squared residuals
$\sum\omega r^2$ (SSR) versus parameters using Wilson's {\bf LC}
code in B filter.}
\end{figure}

\begin{figure}[h]
  \epsfxsize=11cm
  \centerline{
\epsffile{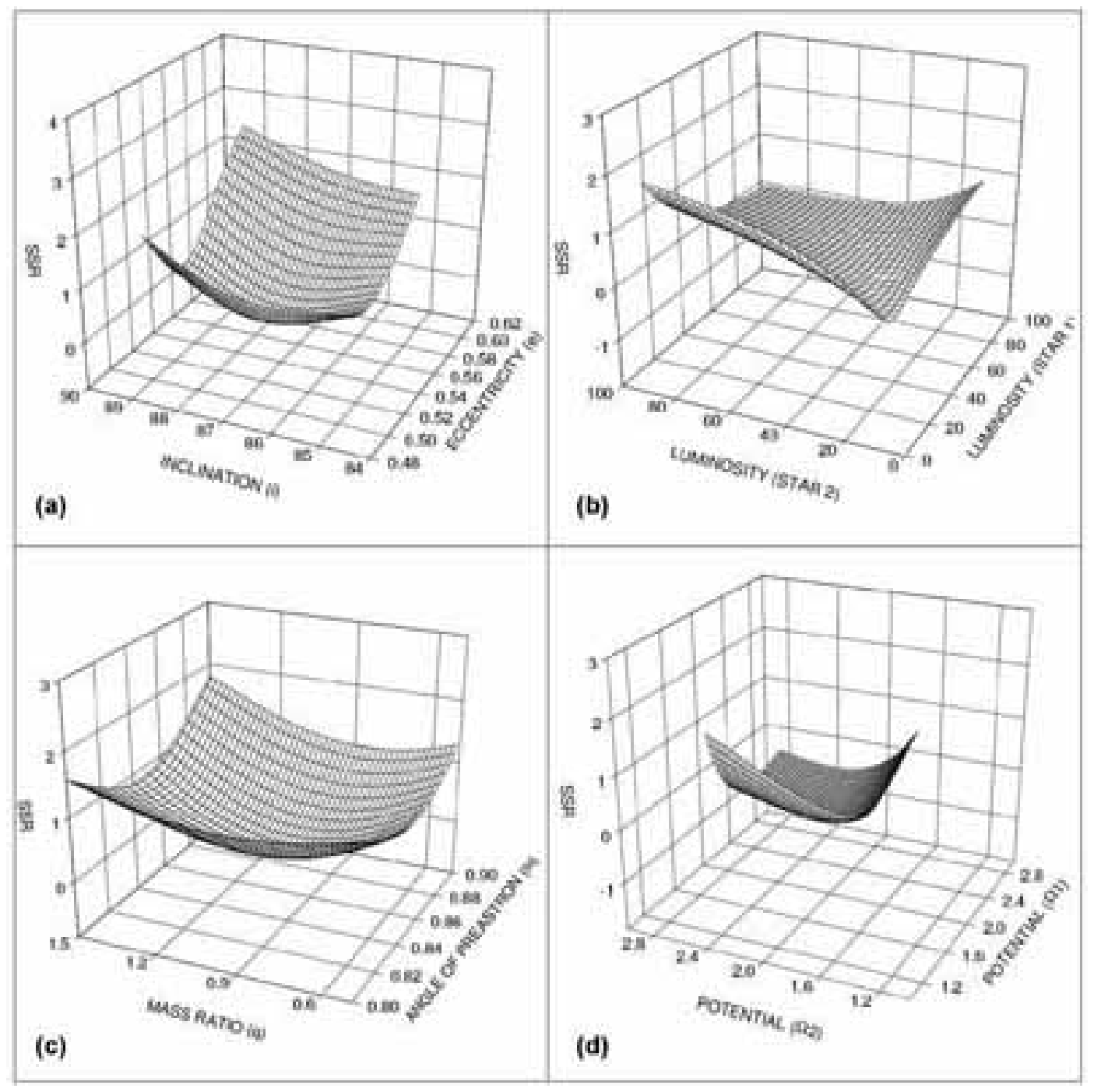} } \caption{The weighted sum of squared residuals
$\sum\omega r^2$ (SSR) versus parameters using Wilson's {\bf LC}
code in V filter.}
\end{figure}

Figures 8 and 9, show the optimized theoretical light curves
together with the observed light curves in B and V filters. The
theoretical curves correspond to the optimized parameters given in
Table 1.\newpage
\begin{center}
Table 1.\\ Optimized parameters of V1143 Cygni.
\end{center}
\begin{tabular}{lll}\hline
Parameter & Filter B &  Filter V \\ \hline
 $i$ & $87.3 \pm0.1$ & $ 87.1 \pm0.1$ \\
$e$ & $0.536 \pm0.005$ & $ 0.539\pm0.005$ \\
$\omega$ & $0^\circ.855 \pm0.004$ & $0^\circ.855 \pm0.004$ \\
$q(\frac{M_2}{M_1})$& $0.99 \pm0.01$ &  $0.98\pm0.01 $\\
$\Omega_1$ & $17.4\pm0.5$ & $18.0\pm0.5 $\\
$\Omega_2$ & $19.6\pm0.5$ & $20.0\pm0.5$ \\
$\frac{L_2}{L_1}$ &$ 0.89 \pm0.05$ &  $0.82\pm0.05$ \\ \hline
\end{tabular}

Finally, the observed times of minima on July 16 and 18, 2002 are
calculated according to the ephemeris \\
\begin{equation}
Min.I=HJD 2447087.5669+7^{d}.64075095 \times E,   \\
\end{equation}

given by Burns et al. (1996)
 and are tabulated in Tables 2 and 3:

\begin{figure}[h]
  \epsfxsize=8cm
  \centerline{
\epsffile{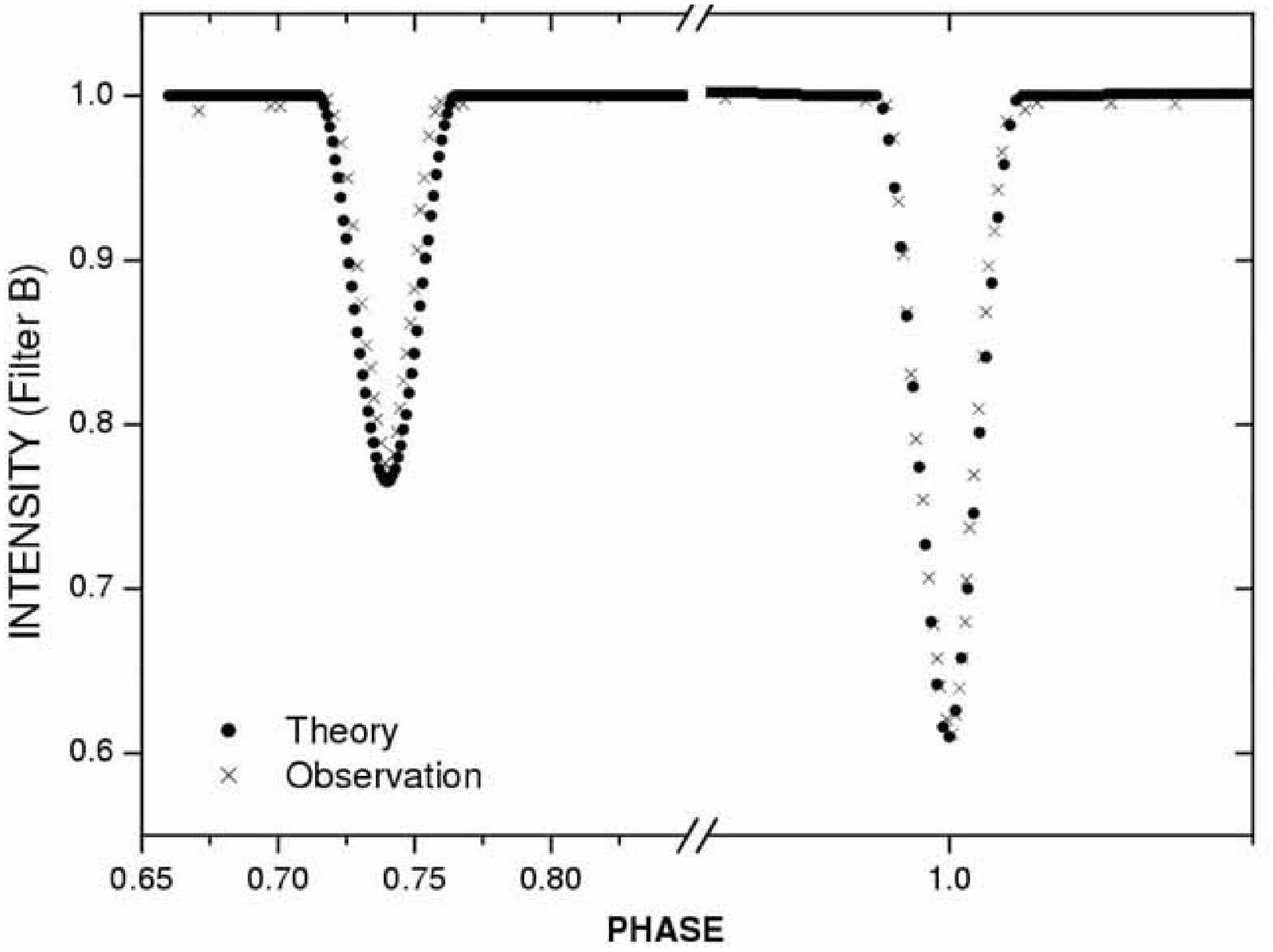} } \caption{Theoretical light curve ({\bf LC}
output)
 in filter B. Points are the observational data.}
\end{figure}

\begin{figure}[h]
  \epsfxsize=8cm
  \centerline{
\epsffile{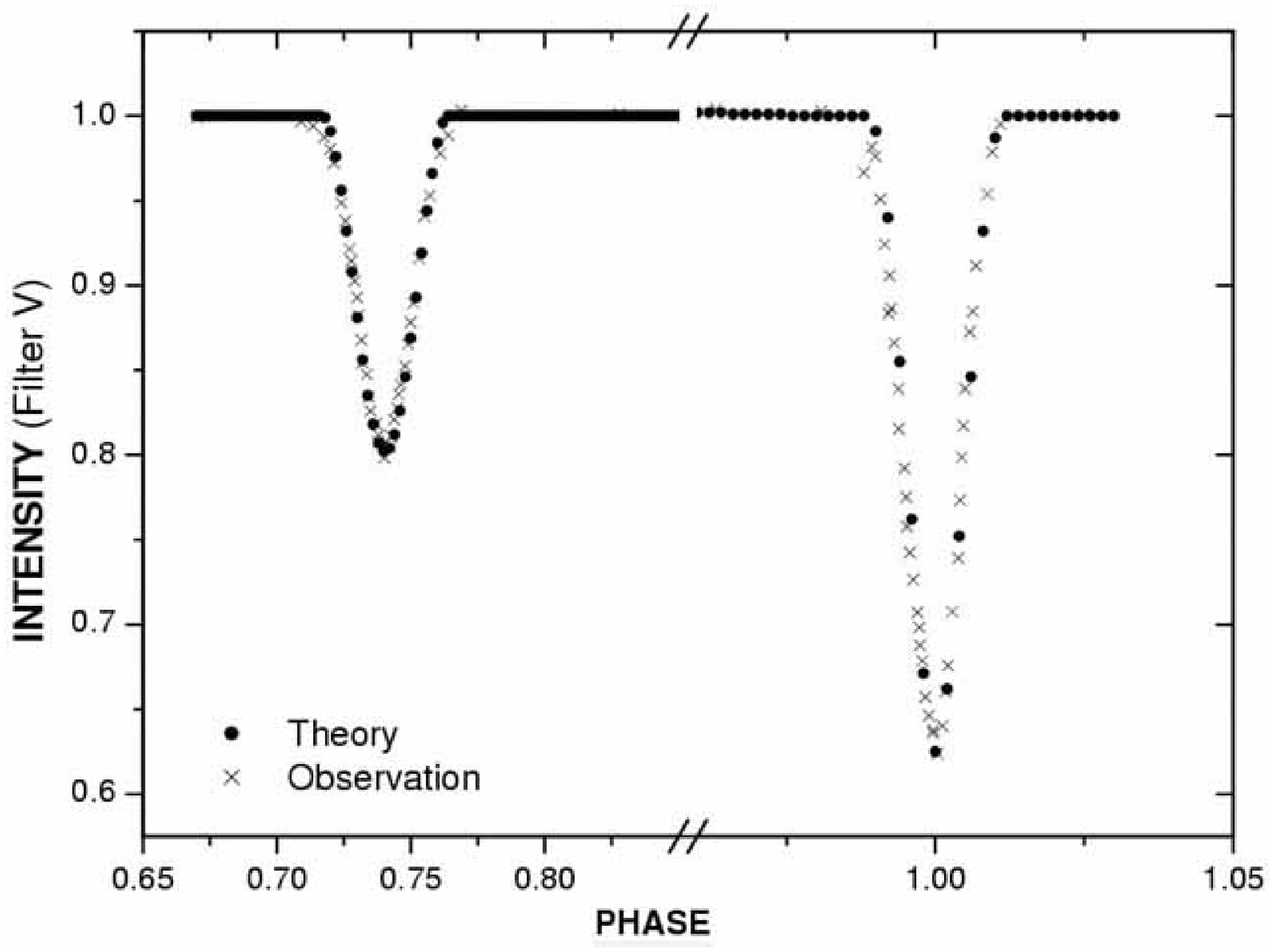} } \caption{Theoretical light curve ({\bf LC}
output)
 in  filter V. Points are the observational data.}
\end{figure}


\section{{\bf Apsidal motion}}
Due to the deep, narrow eclipses of V1143 Cyg and its high eccentricity, the rate of apsidal
motion can be determined exactly by analysis of primary and secondary eclipse timings.
In their paper, Guinan and Maloney (1985) described the procedure that must be
followed to determine the apsidal motion rate from the change in the displacement of the
 secondary minimum from the half point (0.5 phase) according to
\begin{equation}
D=[(t_{2}-t_{1})-0.5{\times}\rm Period].
\end{equation}
D in turn, is related to $\omega$ the longitude of preastron by the formula given
by Sterne (1939):

\begin{equation}
D=\frac{P}{\pi}[\tan^{-1}( \frac{e\cos({\omega})}{{(1-e^2)}^{1/2}}) + \frac{e\cos({\omega})}
{1-{e^2}{\sin^2({\omega})}} {(1-e^2)}^{1/2}],
\end{equation}
where P is the period and e is eccentricity. The observed photoelectric times of primary
 and secondary minima from 1969 to 2002
are collected in Tables 2 and 3. As you can see from Table 4, the
slow decrease in $D$ is due to the advance of the line of apsides
of the orbit. By computing the slope of a line fitted to all of
the secondary minima given in Table 3, we determined an observed
rate of apsidal motion of
$\dot{\omega}_{obs}=3^{\circ}.72/100^{yr}\pm 0.37/100^{yr}$ (see
Figure 10).
\newpage
\begin{center}
 Table 2.\\
 The photoelectric times of primary minima for V1143 Cyg. The O-Cs are computed according to equation 7.\\

\begin{tabular}{lcrl}\hline

H.JD. & O-C(day) & Epoch & Reference\\
(2400000.+) &    &       &         \\ \hline
39339.616   &    -0.22773 & -376 &   Snowden and Koch (1969)$^1$ \\
39385.6881    &  -0.00011 & -370     &   Snowden and Koch (1969)\\
40837.4313    &   0.00017 & -180     &   Battistini et al. (1973) \\
41135.4208  &     0.00033 & -141 &   Battistini et al. (1973)\\
42212.7651     & -0.00142    & 0     &   Koch (1977) \\
42617.727   &     0.00061  &  53 &   Koch (1977)\\
43305.3943  &     0.00021  & 143 &   Guinan et al. (1987) \\
45253.7858     & -0.00008  & 398     &   Gimenez and Margrave (1985)\\
47087.5669  &     0.00049 &  638 &   Burns at al. (1996)\\
48019.73800    & -0.00016 &  760     &   Caton and Burns (1993) \\
49234.6144     & -0.00336 &  919     &   Lacy and Fox (1994) \\
51771.34410    & -0.00338 & 1251     &   Dariush et al. (2001) \\
52474.29443    & -0.00225 & 1343     &  Dariush et al. (2003)\\
\hline


\end{tabular}
\end{center}
\begin{verbatim}
  1. This minimum is observed spectroscopically and because of its
     unusually high O-C, it is not shown in the O-C diagram.
  2. This minimum was published as  2439385.6831 by Hamme and Wilson
     (1984) which seems to be misprint.
\end{verbatim}

\newpage
\begin{center}
 Table 3.\\
 The photoelectric times of secondary minima for V1143 Cyg, together with\\
 the computed values of $D$ and $\omega$,  using equation 6. The O-Cs
are computed according to equation 8.
\begin{tabular}{lcrllll}\hline

H.JD.  & O-C(day) & Epoch &    D   &   $\omega$ &      Reference\\
 (2400000.+)     &          &       &        &     &               \\ \hline
38932.932    &  -0.00431 &  -430 &  1.8685 & 47.97  &   Snowden and Koch (1969) \\
38978.7807   &   0.00000 &  -424 &  1.8727 & 47.83  &   Snowden and Koch (1969)\\
42615.75     &  -0.01896 &    52 &  1.8439 & 48.82  &   Koch (1977)$^1$\\
43066.575    &   0.00286 &   111 &  1.8646 & 48.11  &   Koch (1977)\\
44487.7482   &  -0.00002 &   297 &  1.8579 & 48.34  &   Gimenez and Margrave (1985)\\
47085.5910   &  -0.00598 &   637 &  1.8449 & 48.78  &   Burns at al. (1996)\\
51792.28645  &  -0.00122 &  1253 &  1.8370 & 49.05  &   Dariush et al. (2001)\\
52472.30445  &  -0.00834 &  1342 &  1.8281 & 49.35  &    Dariush
et al. (2003)   \\ \hline

\end{tabular}
\end{center}

\begin{verbatim}
  1.Due to its unusually high O-C, it is not included in determination of
    the observed rate of apsidal motion.
\end{verbatim}

\begin{figure}[h]
  \epsfxsize=8cm
  \centerline{
\epsffile{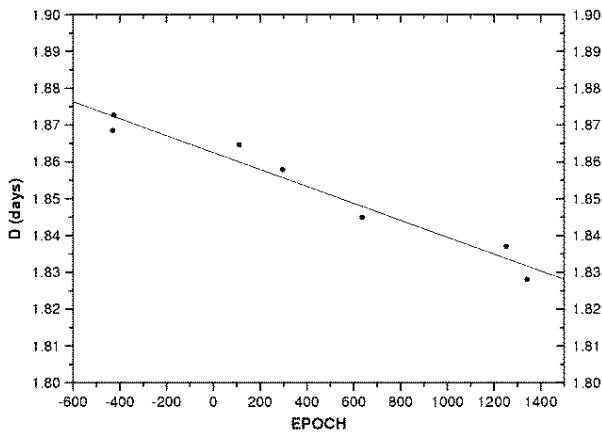} } \caption{The displacement of the secondary
minima given in Table 3,
 from the 0.5 phase versus epoch. The  observed rate of apsidal motion ($\dot{\omega}_{obs}$) can be
 calculated for the slope of this curve.}
\end{figure}

The residuals given in Tables 2 and 3 are calculated for each of
the minima according to the ephemeris given by Gimenez and
Margrave (1985)
\begin{equation}
Min.I =HJD 2442212.76652 + 7^{d}.64075217 \times E,
\end{equation}
and
\begin{equation}
Min.II=HJD 2442218.45092 + 7^{d}.64073165  \times E.
\end{equation}
The computed (O-C)s versus Julian day is plotted in Figure 11 for
both primary and secondary minima. This diagram shows no changes
in period within the 0.01 days.

\begin{figure}[h]
  \epsfxsize=8cm
  \centerline{
\epsffile{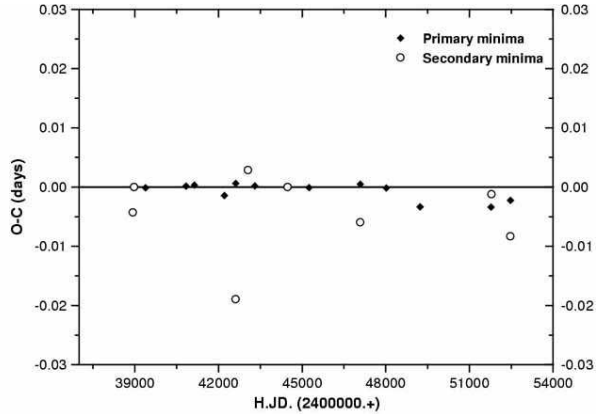} } \caption{The O-C diagram in days for all the
primary and the secondary
 minima tabulated in Tables 2 and 3.}
\end{figure}


\section{{\bf Results and discussion}}
Table 4 contains independent determinations of the observed
apsidal motion of V1143 Cyg together with  the corresponding
period of apsidal revolution $U$. The relation between the apsidal
motion period $U$ and the observed rate of apsidal motion
 ($\dot{\omega}_{obs}$) has a simple form
\begin{equation}
U=\frac{360P}{\dot{\omega}_{obs}},
\end{equation}
where $\dot{\omega}_{obs}$ is expressed in degrees per cycle and
$P$ is the anomalistic period expressed in days. To determine a
more accurate value for $\dot{\omega}_{obs}$ we need more accurate
timings of secondary minima. From Table 4, it seems that expanding
our observational baseline, may decrease the discrepancy between
$\dot{\omega}_{obs}$ and $\dot{\omega}_{theo}$. Figure 12 shows
$D$ as a function of $\omega$ using equation 6 for V1143 Cygni and
DI Herculis. In the case of DI Herculis
 (P=10.550 days, e=0.49)
the discrepancy is much larger than that of V1143 Cygni. For this system the observed
rate of apsidal motion is about one-third of the theoretical one which is due to
the classical and relativistic contribution (Claret, 1998; Dariush and Riazi, 2003).
 The dots in Figure 12, represent our
observational period for the V1143 Cyg and DI Herculis apsidal
revolution. It is clear that up to now, our observational baseline
covers only a very small fraction of the period of apsidal
revolution $U$. In the case of DI Herculis, the presence of a
third body is a plausible  explanation for the  discrepancy
between $\dot{\omega}_{obs}$ and $\dot{\omega}_{theo}$, but until
 no further observational evidence has been reported this
 possibility. Our results for V1143 Cyg supporting $3^\circ.72\pm 0.37/100 yr$
is slightly closer to the theoretical apsidal motion rate
($4^\circ.25\pm 0.72/ 100yr$), computed with previous studies of
this system
\begin{figure}[h]
  \epsfxsize=8cm
  \centerline{
\epsffile{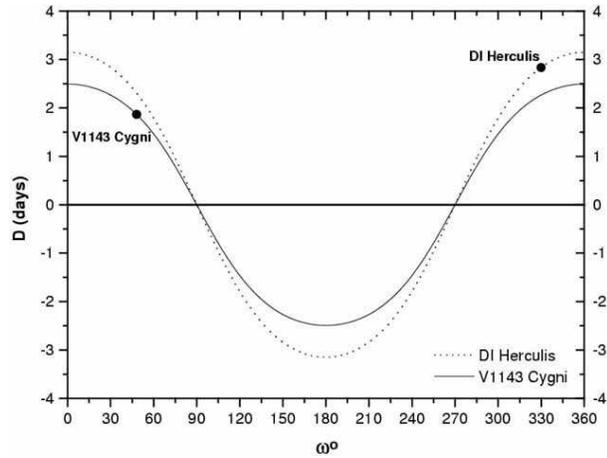} } \caption{ $D$ as a function of $\omega$ using
equation 6 for V1143 Cygni and DI Herculis.}
\end{figure}

\begin{center}
Table 4.\\
Determined rate of apsidal motion for V1143 Cygni
\begin{tabular}{llrl}\hline
${\dot\omega_{obs}}^\circ$/100$^{yr}$ & $\pm$Error& U$^{yr}$ & Source \\ \hline
3.49    & $\pm 0.38$ & 10320  & Khaliullin (1983) \\
3.36    & $\pm 0.19$ & 10710  & Gimenez and Margrave (1985)\\
3.52    & $\pm 0.72$ & 10230  & Burns et al. (1996)\\
3.72    & $\pm 0.37$ &  9680  & Present study$^1$\\ \hline
\end{tabular}
\end{center}

\begin{verbatim}
  1.This value is computed from all the  secondary  minima presented
    in Table 4  except  2442615.75  of  Koch  (1977). Including this
    minimum, it is changed to 3.45 degree/100yr +/- 0.65 degree/100yr.

\end{verbatim}

\begin{center}
{\bf Acknowledgments}\\
\end{center}
We would like to thanks Mr. Mehdi Nazem for his help during the
observations.
\newpage
\begin{center}
{\bf References:}\\
\end{center}
Andersen, J., Garcia, J. M., Gimenez, A., and Nordstr$\ddot o$m, B., 1987,
{\it Astron.Astrophys.}, {\bf 174}, 107.\\
Battistini, P., Bonifazi, A., and Guarnieri, A., 1973, {\it IBVS} {\bf 817}.\\
Burns, J. F., Guinan, E. F., and Marshall, J. J., 1996, {\it IBVS} {\bf 4363}.\\
Caton, D. B., and Burns, W. C., 1993, {\it IBVS} {\bf 3900}.\\
Claret, A. 1998, {\it Astron.Astrophys.}, {\bf 330}, 533.\\
Claret, A., and Willems, B., 2002, {\it Astron.Astrophys.}, {\bf 388}, 518.\\
Dariush, A., Afroozeh, A., Riazi, N., 2001, {\it IBVS} {\bf 5136}.\\
Dariush, A., and Riazi, N., 2003, {\it Astrophys.Space.Sci.}, {\bf 283}, 253.\\
Dariush, A., Zabihinpoor, S. M., Bagheri, M. R., Jafarzadeh, Sh.,
Mosleh, M., and Riazi, N., 2003, {\it IBVS} {\bf 5456}.\\
Gimenez, A., 1985, {\it Astrophys.J.}, {\bf 297}, 405.\\
Gimenez, A., and Margrave, T. E., 1985, {\it Astron.J.}, {\bf 90(2)}, 358.\\
Guinan, E.F. and Maloney, F.P., 1985,{\it Astron.J.}, {\bf 90}, 1519.\\
Guinan, E. F., Najafi, I., Zamani-Noor, F., and Boyd, P. T., 1987, {\it IBVS} {\bf 3070}.\\
Hamme, W. V., and Wilson, R. E., 1984, {\it Astron.Astrophys.}, {\bf 141}, 1.\\
Khaliullin, Kh. F., 1983, {\it Astron.Trisk}, No 1262.\\
Koch, R. H., 1977, {\it Astron.J.},{\bf 82}, 653.\\
Lacy, C.H. and Fox, G.W., 1994, {\it IBVS} {\bf 4009}.\\
Semeniuk, I., 1968, {\it Acta.Astron.}, {\bf 18}, 1.\\
Snowden, M. S., and Koch, R. H., 1969, {\it Astrophys.J.},{\bf 156}.\\
Sterne, T.E., 1939(a), {\it Mon.Not.R.Astron.Soc.}, {\bf 99}, 451.\\
Sterne, T.E., 1939(b), {\it Mon.Not.R.Astron.Soc.}, {\bf 99}, 662.\\
Wilson, R.E., 1979, {\it Astrophys.J.}, {\bf 234}, 1054.\\
Wilson, R.E., 1990, {\it Astrophys.J.}, {\bf 356}, 613.\\
Wilson, R.E., 1993, in {\it New Frontiers in Binary Star
Research}, ed. K. C. Leung and
Nha, I.S., A.S.P. Conf. Ser., {\bf 38}, 91.\\

\end{document}